\documentclass[12 pt]{article}
\usepackage{amsfonts}
\usepackage{amsmath}
\usepackage{geometry}
\usepackage{cite}
\begin{document}

\title{Properties of shape-invariant tridiagonal Hamiltonians}
\author{ Hashim A. Yamani$^*$ and Zouha\"{\i}r Mouayn$^\ddagger$ 
}
\maketitle
\begin{center}
$^*$Mabuth-2414, Medina 42362-6959, Saudi Arabia\\
$^\ddagger$Department of Mathematics, Faculty of
Sciences and Technics (M'Ghila)\\P.O. Box. 523, B\'{e}ni Mellal, Morocco.
\end{center}
\begin{abstract}
It has been established that a positive semi-definite Hamiltonian,$H$, that
has a tridiagonal matrix representation in a basis set, allows a definition
of forward (and backward) shift operators that can be used to define the
matrix representation of the supersymmetric partner Hamiltonian $H^{\left(
+\right) \text{\ }}$ in the same basis. \ We show that if, additionally, the
Hamiltonian has a shape invariant property, the matrix elements of the
Hamiltonian are related in a such a way that the energy spectrum is known in
terms of these elements. It is also possible to determine the matrix
elements of the hierarchy of super-symmetric partner Hamiltonians.
Additionally, we derive the coherent states associated with this type of
Hamiltonians and illustrate our results with examples from well-studied
shape-invariant Hamiltonians that also has tridiagonal matrix representation.

\textbf{PACS}: 02.60 Jh (or generally, 02.60.-x), 02.07.-c

\textbf{Keywords}:Supersymmetry, Shape-invariant potentials, Superpotential,
Raising and lowering operators, Coherent states.
\end{abstract}

\section{Introduction}

With the idea of supersymmetric quantum mechanics (SUSY) the concept of
shape invariance was put forward by Gendenshtein \cite{Gend}. The
condition of shape invariance requires from the supersymmetric partner potentials a
condition of type $V_{\eta }^{\left( +\right) }\left( x\right) =V_{f\left(
\eta \right) }\left( x\right) +R\left( \eta \right) $ where $f\left( \eta
\right) $ is a function of $\eta $ and $R\left( \eta \right) $ is
independent of $x.$ This is equivalent to the operator relation $H^{\left(
+\right) }\left( \eta \right) =H\left( f\left( \eta \right) \right) +R\left(
\eta \right) $ which  is also an integrability condition \cite{Gend} that has been proved sufficient to get exact results. Indeed, SUSY and the
shape-invariance condition provide an algebraic procedure to determine the
entire spectrum of solvable quantum systems, without any need to solve a
differential equation. Almost exactly one-dimensional potential problems
encountered in quantum mechanics are shape invariant where the parameters
are related by a translation $\gamma =f\left( \eta \right) =\eta +\delta $ \cite{Cooper1995}.
Actually, with such potentials is associated the Lie algebra $so\left(
2,1\right) $ which was used to recover the spectrum of the model. For a review of supersymmetry, shape invariance and exactly solvable potential see \cite{RAU}.

Traditionally, the literature of SUSY quantum mechanics is cast in terms of
the system Hamiltonian's representation in configuration space \cite{Cooper2001}. One of the benefits of this representation is to deduce the
explicit form, in configuration space, of the partner potential and the
so-called superpotential. Alternatively, working with the matrix
representation of a Hamiltonian $H$ in a given basis set $\left( \mid
\phi \rangle \right) _{n=0}^{\infty }$ has proved to be a viable description
of the physical system ever since the beginning years of the development of
quantum mechanics. It is by now an established fact that many calculation
tools in quantum mechanics employ the language of matrix representation of
physical operators in a basis \cite{MS}. The authors have already shown \cite{MY14 , MY16} that if the matrix representation of the Hamiltonian in the basis
is  tridiagonal,%
\begin{equation}
H_{n,m}=\langle \phi _{n}\mid H\mid \phi _{m}\rangle =b_{n-1}\delta
_{n,m+1}+a_{n}\delta _{n,m}+b_{n}\delta _{n,m-1}  \tag{1.1}
\end{equation}
a supersymmetric partner Hamiltonian $H^{\left( +\right) }$ also has a
tridiagonal representation in the same basis,%
\begin{equation}
H_{n,m}^{\left( +\right) }=\langle \phi _{n}\mid H^{\left( +\right) }\mid
\phi _{m}\rangle =b_{n-1}^{\left( +\right) }\delta _{n,m+1}+a_{n}^{\left(
+\right) }\delta _{n,m}+b_{n}^{\left( +\right) }\delta _{n,m-1},  \tag{1.2}
\end{equation}%
with the coefficients related as%
\begin{equation}
a_{n}= c_{n}^{2}+ d_{n}
^{2},b_{n}=c_{n}d_{n+1}  \tag{1.3}
\end{equation}%
\begin{equation}
a_{n}^{\left( +\right) }= c_{n} ^{2}+
d_{n+1}^{2},b_{n}^{\left( +\right) }=c_{n+1}d_{n+1}. 
\tag{1.4}
\end{equation}%
Here the two sets of coefficients $\left( c_{n}\right) $ and $\left(
d_{n}\right) $ figure in the definition of the forward-shift operator $A$
through specifying its action on a basis vector as%
\begin{equation}
A\mid \phi _{m}\rangle =c_{m}\mid \phi _{m}\rangle +d_{m}\mid \phi
_{m-1}\rangle  \tag{1.5}
\end{equation}%
This means that the action of its adjoint $A^{\dagger }$ on a basis is now 
\begin{equation}
A^{\dagger }\mid \phi _{m}\rangle =c_{m}\mid \phi _{m}\rangle +d_{m+1}\mid
\phi _{m+1}\rangle.  \tag{1.6}
\end{equation}%
The Hamiltonian and its supersymmetric partner are related to these
operators as 
\begin{equation}
H=A^{\dagger }A,\text{ \ }H=AA^{\dagger }.  \tag{1.7}
\end{equation}%
More explicitly, the parameters $\left( c_{n}\right) $ and $\left(
d_{n}\right) $ can be calculated recursively if the tridiagonal matrix
elements of the Hamiltonian are known, or alternatively from the relations 
\begin{equation}
c_{n}^{2}=-b_{n}\frac{P_{n+1}\left( \varepsilon _{0}\right) }{P_{n}\left(
\varepsilon _{0}\right) },\qquad d_{n+1}^{2}=-b_{n}\frac{P_{n}\left( \varepsilon
_{0}\right) }{P_{n+1}\left( \varepsilon _{0}\right) }  \tag{1.8}
\end{equation}%
where $\varepsilon _{0}$ is the energy of the ground state. Here, the set $%
\left( P_{n}\left( \varepsilon \right) \right) $ is a solution of three-term
recursion relation 
\begin{equation}
\varepsilon P_{n}\left( \varepsilon \right) =b_{n-1}P_{n-1}\left(
\varepsilon \right) +a_{n}P_{n}\left( \varepsilon \right)
+b_{n}P_{n+1}\left( \varepsilon \right) .  \tag{1.9}
\end{equation}%
In this paper, we intent to explore the implications of shape invariance and
find the additional properties that the above basic parameters satisfy and
which in turn enable us to find solution to the physical quantities
associated with the system Hamiltonian. In section 2, we explore the form of various parameters list
above and particular the set $\left\{ c_{n},d_{n}\right\} $ which plays
a fundamental role in the descriptions. We show that we may define a
modified version of the operators $A$ and $A^{\dagger }$ which can be
interpreted essentially as lowering and raising operators. This allows us to
specify the full energy spectrum of the system using only the parameters $\left\{
c_{0},d_{1}\right\} .$ This also leads easily to the determination of the
supersymmetric partner potential. \ We also show that the shape-invariance
property allows the complete specification of the set of $c_{n}-$like and
the $d_{n}-$like parameters associated with the hierarchy of supersymmetric
partner Hamiltonians. Although in our setup the concept of superpotential
plays no role in describing the system, we, nonetheless, write down what the
form of a quantity that has all the properties of the familiar
superpotential. In section 3, we explicitly construct the coherent states
associated with a shape-invariant tridiagonal Hamiltonian and show by
examples that it possesses the expected properties. Section 4 is devoted to
some concluding remarks. Throughout, we illustrate our results using cases
previously studied leaving details to the appendices.

\section{Implications of shape-invariance}

\subsection{Properties of the matrix elements of shape-invariant Hamiltonians%
}

A given positive semi-definite Hamiltonian $H$ (with the ground state $\varepsilon_0$ set to zero) satisfying a shape-invariance
property is related to its supersymmetric partner Hamiltonian $H^{\left(
+\right) }$ via the relation 
\begin{equation}
H^{\left( +\right) }\left( \eta \right) =H\left( \eta +\delta \right)
+R\left( \eta \right)  \tag{2.1.0}
\end{equation}%
where the parameters $\eta ,\delta $ and $R$ are property of the given
Hamiltonian. Since the spectrum of $H^{\left( +\right) }$ is shifted by an
amount of $\varepsilon _{1}$ compared to the spectrum of $H,$ we must have $%
R\left( \eta \right) =\varepsilon _{1}\left( \eta \right) .$ Furthermore, if 
$H$ has a tridiagonal representation in a basis as in Eq.$\left( 1.1\right) $%
, then so does $H^{(+)}$ as in Eq.$\left( 1.2\right) .$ Hence the above
symmetry relation translates into the following connections among their
matrix elements, 
\begin{equation}
a_{n}^{\left( +\right) }\left( \eta \right) =a_{n}\left( \eta +\delta
\right) +\varepsilon _{1}\left( \eta \right),  \tag{2.1.1}
\end{equation}%
\begin{equation}
b_{n}^{\left( +\right) }\left( \eta \right) =b_{n}\left( \eta +\delta \right).
\tag{2.1.2}
\end{equation}%
From the relations $\left( 1.3\right) -\left( 1.4\right) $ of the parameters $%
\left( a_{n}^{\left( +\right) },b_{n}^{\left( +\right) }\right) $ and $%
\left( a_{n},b_{n}\right) $ to the basic coefficients $\left(
c_{n},b_{n}\right) $, the above relations means that

\begin{equation}
c_{n}^{2}\left( \eta \right) +d_{n+1}^{2}\left( \eta \right)
=c_{n}^{2}\left( \eta +\delta \right) +d_{n}^{2}\left( \eta +\delta \right)
+\varepsilon _{1}\left( \eta \right)  \tag{2.1.3}
\end{equation}%
\begin{equation}
c_{n+1}\left( \eta \right) d_{n+1}\left( \eta \right) =c_{n}\left( \eta
+\delta \right) d_{n+1}\left( \eta +\delta \right)  \tag{2.1.4}
\end{equation}%
We now postulate that the coefficients $\left( d_{n}\right) $ are
independent of the parameter $\eta .$ This turns out to be the case for many
of the known shape-invariant Hamiltonians such as Harmonic oscillator and
the Morse Hamiltonians. In that case, $d_{n}\left( \eta \right) =d_{n}\left(
\eta +\delta \right) =d_{n}.$ This gives $c_{n+1}\left( \eta \right)
=c_{n}(\eta +\delta ).$ Putting the above relations together, we get 
\begin{equation}
\varepsilon _{1}\left( \eta \right) =\left( c_{n}^{2}\left( \eta \right)
-c_{n+1}^{2}\left( \eta \right) \right) +\left( d_{n+1}^{2}-d_{n}^{2}\right)
\tag{2.1.5}
\end{equation}%
We make two remarks regarding the above important result. First, to
calculate $\varepsilon _{1}\left( \eta \right) $ there is no need to
identify explicitly what the parameter $\eta $ is. The second is that this
result is independent of $n$. Specifically, if we set $n=0,$ we have the
simple yet powerful relation%
\begin{equation}
\varepsilon _{1}\left( \eta \right) =\left( c_{0}^{2}\left( \eta \right)
-c_{1}^{2}\left( \eta \right) \right) +d_{1}^{2}  \tag{2.1.6}
\end{equation}%
giving $\varepsilon _{1}\left( \eta \right) $ in terms of only the three
parameters $\left\{ c_{0}\left( \eta \right) ,c_{1}\left( \eta \right)
,d_{1}\right\} .$

As is suggestive in the symmetry property Eq. $\left( 2.1.0\right) $, we
will find it convenient to utilize the translation operator $T:=e^{-\delta 
\frac{\partial }{\partial \eta }}.$ In fact, if $K\left( \eta \right) $ is
any operator, then $T^{\dag }K\left( \eta \right) T=K\left( \eta +\delta
\right) $. Here, of course, $T^{\dag }=e^{\delta \frac{\partial }{\partial
\eta }}.$ We choose to integrate this property in the basic setup by
replacing the operator $A$ by the operator $B=TA$ and its adjoint $B^{\dag
}=A^{\dag }T^{\dag }$. Remarkably, we still have $H\left( \eta \right)
=A^{\dag }A=B^{\dag }B.$ On the other hand, we have a new version of the
supersymmetric partner Hamiltonian, namely $\overline{H}^{\left( +\right)
}\left( \eta \right) =BB^{\dag }.$ It is related to $H^{\left( +\right)
}\left( \eta \right) $ simply as 
\begin{equation}
\overline{H}^{\left( +\right) }\left( \eta \right) =BB^{\dag }=TAA^{\dag
}T^{\dag }=H^{\left( +\right) }\left( \eta -\delta \right).  \tag{2.1.7}
\end{equation}%
Because of this relation, it is clear that $\overline{H}^{\left( +\right)
}\left( \eta \right) $ also has a tridiagonal matrix representation in the
same basis, namely%
\begin{equation*}
\overline{H}^{\left( +\right) }\left( \eta \right) \mid \phi _{n}\rangle
=c_{n}\left( \eta -\delta \right) d_{n}\mid \phi _{n-1}\rangle +\left(
c_{n}^{2}\left( \eta -\delta \right) +d_{n+1}^{2}\right) \mid \phi
_{n}\rangle +c_{n+1}\left( \eta -\delta \right) d_{n+1}\mid \phi
_{n+1}\rangle . \tag{2.1.8}
\end{equation*}%
Now comparing $BB^{\dag }\mid \phi _{n}\rangle $ and $B^{\dag }B\mid \phi
_{n}\rangle $ \ while utilizing the symmetry properties $d_{n}\left( \eta
\right) =d_{n}\left( \eta +\delta \right) =d_{n}$ and $c_{n+1}\left( \eta
\right) =c_{n}(\eta +\delta ),$ we get 
\begin{equation}
\left[ B,B^{\dag }\right] \mid \phi _{n}\rangle =\left( \left[
c_{n-1}^{2}\left( \eta \right) +d_{n+1}^{2}\right] -\left[ c_{n}^{2}\left(
\eta \right) +d_{n}^{2}\right] \right) \mid \phi _{n}\rangle .  \tag{2.1.9}
\end{equation}%
It is not hard to recognize that the quantity on the right-hand side is just 
$\varepsilon _{1}\left( \eta -\delta \right) \mid \phi _{n}\rangle .$ We
thus obtain the major result that 
\begin{equation*}
\left[ B,B^{\dag }\right] =\varepsilon _{1}\left( \eta -\delta \right). \tag{2.1.10}
\end{equation*}%
This commutation relation suggests that the operators $B^{\dag }$ and $B$
may be considered as raising and lowering operators provided their
noncommutativity with $\varepsilon _{1}\left( \eta \right) $ is taken into
account. The noncommutativity of the operators $B^{\dag }$ and $B$ with any
function $f\left( \eta \right) $ is given by the following two very
important relations $f\left( \eta \right) B^{\dag }=B^{\dag }f\left( \eta
-\delta \right) $ and $f\left( \eta \right) B=Bf\left( \eta +\delta \right)
. $

We list in Table 1 results that apply to three shape-invariant Hamiltonians
having tridiagonal matrix representation, namely the $l$-th partial wave
Kinetic energy, Harmonic oscillator and Morse Hamiltonians.

\subsection{The operator $B^{\dag }$ as a raising operator}

In this subsection, we clarify the sense of $B^{\dag }$ behaving as a
raising operator. In fact, we will show that%
\begin{equation}
\mid \varepsilon _{m+1}\left( \eta \right) \rangle =\frac{1}{\sqrt{%
\sum\limits_{k=0}^{m}\varepsilon _{1}\left( \eta +k\delta \right) }}B^{\dag }%
\frac{1}{\sqrt{\sum\limits_{k=0}^{m-1}\varepsilon _{1}\left( \eta +k\delta
\right) }}B^{\dag }...\frac{1}{\sqrt{\varepsilon _{1}\left( \eta \right)
+\varepsilon _{1}\left( \eta +\delta \right) }}B^{\dag }\frac{1}{\sqrt{%
\varepsilon _{1}\left( \eta \right) }}B^{\dag }\mid \varepsilon _{0}\left(
\eta \right) \rangle  \tag{2.2.0}
\end{equation}%
The proof can be proceed by induction (see Appendix A). An iterative
version of the relation is%
\begin{equation}
\mid \varepsilon _{m+1}\left( \eta \right) \rangle =\frac{1}{\sqrt{%
\sum\limits_{k=0}^{m}\varepsilon _{1}\left( \eta +k\delta \right) }}B^{\dag
}\mid \varepsilon _{m}\left( \eta \right) \rangle .  \tag{2.2.1}
\end{equation}%
The latter one shows even more the sense in which $B^{\dag }$ is considered
as a raising operator since it can generate the higher energy eigenstates not
only from the ground state as in Eq.(2.2.0) but alternatively from the level
just below it. This property will be exploited to construct coherent states
using the displacement operator $D\left( z\right) =\exp \left( zB^{\dag
}-z^{\ast }B\right) $ as we will show in Section 3. As an example, we carry
out explicitly in Appendix B a calculation that shows that the operator $B$
acts as a lowering operator for the harmonic oscillator Hamiltonian.

\subsection{The energy spectrum}

It is remarkable how the shape-invariance together with minimal extra
information leads to full characterization of the system's energy spectrum.
To see this, we note that from Eq.(2.1.0), 
\begin{equation*}
H^{\left( +\right) }\left( \eta -\delta \right) =H\left( \eta \right)
+\varepsilon _{1}\left( \eta -\delta \right). \tag{2.3.0}
\end{equation*}%
Since the operators $B^{\dag }$ and $B$ are treated as raising and lowering
\ operators, the Hamiltonians share the same spectrum except for the ground
state. So, if $H\left( \eta \right) \mid \varepsilon _{m}\rangle
=\varepsilon _{m}\left( \eta \right) \mid \varepsilon _{m}\rangle ,$ then
the above equation gives 
\begin{equation*}
H^{\left( +\right) }\left( \eta -\delta \right) \mid \varepsilon _{m}\rangle
=\left[ H\left( \eta \right) +\varepsilon _{1}\left( \eta -\delta \right) %
\right] \mid \varepsilon _{m}\rangle \tag{2.3.1}
\end{equation*}

\begin{equation*}
\varepsilon _{m}^{\left( +\right) }\left( \eta -\delta \right) \mid
\varepsilon _{m}\rangle =\left[ \varepsilon _{m}\left( \eta \right)
+\varepsilon _{1}\left( \eta -\delta \right) \right] \mid \varepsilon
_{m}\rangle. \tag{2.3.2}
\end{equation*}%
Hence, 
\begin{equation*}
\varepsilon _{m}^{\left( +\right) }\left( \eta -\delta \right) =\left[
\varepsilon _{m}\left( \eta \right) +\varepsilon _{1}\left( \eta -\delta
\right) \right] \tag{2.3.3}
\end{equation*}%
or 
\begin{equation*}
\varepsilon _{m}^{\left( +\right) }\left( \eta \right) =\left[ \varepsilon
_{m}\left( \eta +\delta \right) +\varepsilon _{1}\left( \eta \right) \right] \tag{2.3.4}
\end{equation*}%
On the other hand, if we write $\varepsilon _{m}^{\left( +\right) }\left(
\eta \right) =\varepsilon _{m+1}$ then an interesting new result emerges~for
tridiagonal Hamiltonians satisfying the shape invariance property, namely,

\begin{equation*}
\varepsilon _{m+1}\left( \eta \right) =\varepsilon _{m}\left( \eta +\delta
\right) +\varepsilon _{1}\left( \eta \right)
=\sum\limits_{k=0}^{m}\varepsilon _{1}\left( \eta +k\delta \right). \tag{2.3.5}
\end{equation*}%
Since we can derive $\varepsilon _{m}\left( \eta \right) $ and $\varepsilon
_{m}^{\left( +\right) }\left( \eta \right) $ from $\varepsilon _{1}\left(
\eta \right) ,$ and since $\varepsilon _{1}\left( \eta \right) $ can be
derived from the three parameters $\left\{ c_{0},c_{1,}d_{1}\right\} $ as
per Eq.(2.1.6), the system energy spectrum can essentially be derived from
these parameters as well. Alternatively, if we know $\left\{ \left(
c_{n}\right) _{n=0}^{\infty },d_{1}\right\} ,$ then we can find a very
simple expression for $\varepsilon _{m}\left( \eta \right) $ or $\varepsilon
_{m}^{\left( +\right) }\left( \eta \right) .$ From Eq.(2.1.5)
we have 
\begin{equation*}
\varepsilon _{1}\left( \eta +k\delta \right) =\left[ c_{n}^{2}\left( \eta
+k\delta \right) -c_{n+1}^{2}\left( \eta +k\delta \right) \right] +\left[
d_{n+1}^{2}-d_{n}^{2}\right] .
\end{equation*}%
Also, from Eq.(2.1.3) and Eq.(2.1.4) 
\begin{equation*}
\varepsilon _{1}\left( \eta +k\delta \right) =\left[ c_{n}^{2}\left( \eta
\right) -c_{n+k+1}^{2}\left( \eta \right) \right] +\left[
d_{n+1}^{2}-d_{n}^{2}\right] .
\end{equation*}%
Then from $d_n(\eta)=d_n(\eta+\delta)=d_n$ and $c_{n+1}(\eta)=c_n(\eta+\delta)$ we have%
\begin{equation*}
\varepsilon _{m}^{\left( +\right) }\left( \eta \right) =\varepsilon
_{m+1}\left( \eta \right) =\sum\limits_{k=0}^{m}\varepsilon _{1}\left( \eta
+k\delta \right) =\sum\limits_{k=0}^{m}\left( \left[ c_{m+k}^{2}\left( \eta
\right) -c_{m+k+1}^{2}\left( \eta \right) \right] +\left[
d_{m+1}^{2}-d_{m}^{2}\right] \right)
\end{equation*}%
We now use the telescoping property for the last sum. Specifically, we have%
\begin{equation*}
\varepsilon _{m}^{\left( +\right) }\left( \eta \right) =\varepsilon
_{m+1}\left( \eta \right) =\sum\limits_{k=0}^{m}\varepsilon _{1}\left( \eta
+k\delta \right) =\left[ c_{m+k}^{2}\left( \eta \right) -c_{m+k+1}^{2}\left(
\eta \right) \right] +\left( m+1\right) \left[ d_{m+1}^{2}-d_{m}^{2}\right] .
\end{equation*}%
A simple expression results from specializing to the case $n=0,$%
\begin{equation*}
\varepsilon _{m}^{\left( +\right) }\left( \eta \right) =\varepsilon
_{m+1}\left( \eta \right) =\left[ c_0^{2}\left( \eta \right)
-c_{m+1}^{2}\left( \eta \right) \right] +\left( m+1\right) d_{1}^2. \tag{2.3.6}
\end{equation*}%
As examples, the parameters listed in Table 1 can be used to show how the
above results apply to the listed Hamiltonians.

\subsection{Supersymmetric partner potential}

If the Hamiltonian satisfies the shape-invariance property of Eq.(2.1.0), then
we have in more details 
\begin{equation*}
H_{0}\left( \eta \right) +V^{\left( +\right) }\left( \eta \right)
=H_{0}\left( \eta +\delta \right) +V\left( \eta +\delta \right) +\varepsilon
_{1}\left( \eta \right) \tag{2.4.0}
\end{equation*}%
Since the reference Hamiltonian $H_{0}\left( \eta \right) =-\frac{1}{2}\frac{%
d^{2}}{dx^{2}},$ we must have $H_{0}\left( \eta \right) $ independent of the parameter $\eta $. Hence $H_{0}\left( \eta \right) =H_{0}\left( \eta
+\delta \right) .$ Therefore, the two partner potential are related as 
\begin{equation*}
V^{\left( +\right) }\left( \eta \right) =V\left( \eta +\delta \right)
+\varepsilon _{1}\left( \eta \right) \tag{2.4.1}
\end{equation*}%
We verify this relation for three familiar potential in Table 1.
\subsection{The coefficients $\left( c_{n},d_{n}\right) $ associated with
the hierarchy of supersymmetric shape-invariant Hamiltonians}

For a given positive semi-definite tridiagonal Hamiltonian $H\left( \eta
\right) $ (with the ground state energy $\varepsilon _{0}\left( \eta \right)
=0$), the coefficients $\left( c_{n},d_{n}\right) $ are related to the
coefficients $\left( a_{n},b_{n}\right) $ as in Eq.(1.3), namely, $%
a_{n}=c_{n}^{2}+d_{n}^{2},$ $b_{n}=c_{n}d_{n+1}.$ We then ask the question:
what are the coefficients $\left( c_{n}^{\left( +\right) },d_{n}^{\left(
+\right) }\right) $ that are associated with the partner Hamiltonian $%
H^{\left( +\right) }\left( \eta \right) $, so that we can similarly write%
\begin{equation*}
a_{n}^{\left( +\right) }=\left( c_{n}^{\left( +\right) }\right) ^{2}+\left(
d_{n}^{\left( +\right) }\right) ^{2},\text{ \ }b_{n}^{\left( +\right)
}=c_{n}^{\left( +\right) }d_{n+1}^{\left( +\right)} \,\, ? \tag{2.5.0}
\end{equation*}%
Since we already know that%
\begin{equation*}
a_{n}^{\left( +\right) }=c_{n}^{2}+d_{n}^{2},\text{ \ \ \ }b_{n}^{\left(
+\right) }=c_{n+1}d_{n+1} \tag{2.5.1}
\end{equation*}%
the answer as suggested in \cite{MY14} is $c_{n}^{\left( +\right)
}=d_{n+1}$ and $d_{n}^{\left( +\right) }=c_{n}$. This is a general result.
If we now repeat the same question regarding the spersymmetric partner $%
H^{\left( ++\right) }\left( \eta \right) $ in the next level of hierarchy,
we have to be careful to seek the supersymmetric partner of the Hamiltonian%
\begin{equation}
\widetilde{H}^{\left( +\right) }\left( \eta \right) =H^{\left( +\right)
}\left( \eta \right) -\varepsilon _{1}\left( \eta \right)  \tag{2.5.2}
\end{equation}%
that has the ground state $\widetilde{\varepsilon }_{0}^{\left( +\right) }=0.$ In
general, we do not know the explicit form of the energy $\varepsilon
_{1}\left( \eta \right) .$ However, for the shape invariant Hamiltonian, the
set of parameters $\left\{ c_{0},c_{1},d_{1}\right\} $ is enough to
completely specify the energy spectrum, as has been demonstrated in the
previous subsections. We now show that shape invariance enables us to find
these coefficients associated with the hierarchy of supersymmetric partner
Hamiltonians.

We now consider, as a starting point the Hamiltonian $\widetilde{H}^{\left(
+\right) }\left( \eta \right) $ and seek its super-symmetric partner. From
Eq.(2.5.2) and the explicit form of $\varepsilon _{1}$ as given by Eq.(2.1.6),
we simply write the following two relations.%
\begin{equation*}
\widetilde{a}_{n}^{\left( +\right) }=a_{n}^{\left( +\right) }-\varepsilon
_{1}\left( \eta \right) =\left( c_{n}^{2}+d_{n+1}^{2}\right) -\left(
c_{n}^{2}\left( \eta \right) +c_{n+1}^{2}\left( \eta \right) +\left(
d_{n+1}^{2}-d_{n}^{2}\right) \right) \tag{2.5.3}
\end{equation*}%
\begin{equation*}
=c_{n+1}^{2}\left( \eta \right) +d_{n}^{2}, \tag{2.5.4}
\end{equation*}%
\begin{equation*}
\widetilde{b}_{n}^{\left( +\right) }=b_{n}^{\left( +\right) }=c_{n+1}d_{n+1}. \tag{2.5.5}
\end{equation*}%
It is easy to conclude that the pair $\left( \widetilde{c}_{n}^{\left(
+\right) },\widetilde{d}_{n}^{\left( +\right) }\right) $ is just $(c_{n+1},d_n)$. We can now write the explicit form of the pair $\left(
c_{n}^{\left( ++\right) },d_{n}^{\left( ++\right) }\right) $ associated with 
$H^{\left( ++\right) }\left( \eta \right) $ which is the supersymmetric
partner Hamiltonian to $\widetilde{H}^{\left( +\right) }\left( \eta \right)
. $ That is, $c_{n}^{\left( ++\right) }=d_{n+1}$ and $d_{n}^{\left(
++\right) }=c_{n+1}.$ This immediately leads to following explicit form of
the pair $\left( a_{n}^{\left( ++\right) },b_{n}^{\left( ++\right) }\right) $
as follows: $a_{n}^{\left( ++\right) }=c_{n+1}^{2}+d_{n+1}^{2}$ and $%
b_{n}^{\left( ++\right) }=c_{n+2}d_{n+1}.$ Because of the shape-invariance
property, we now verify that 
\begin{equation*}
H^{\left( ++\right) }\left( \eta \right) =\widetilde{H}^{\left( +\right)
}\left( \eta +\delta \right) +\widetilde{R }\left( \eta \right) \tag{2.5.6}
\end{equation*}%
with $\widetilde{R}\left( \eta \right) =\left( \varepsilon
_{2}-\varepsilon _{1}\right) .$ To do that, we first note that $%
b_{n}^{\left( ++\right) }=\widetilde{b}_{n}^{\left( +\right) }\left( \eta
+\delta \right) .$ Next, we have 
\begin{equation*}
\widetilde{R}\left( \eta \right) =a_{n}^{\left( ++\right) }\left( \eta
\right) -\widetilde{a}_{n}^{\left( +\right) }\left( \eta +\delta \right)
=\left( d_{n+1}^{2}-d_{n}^{2}\right) -\left( c_{n+2}^{2}\left( \eta \right)
-c_{n+1}^{2}\left( \eta \right) \right) =\varepsilon _{1}\left( \eta +\delta
\right)
\end{equation*}%
which is just $\left( \varepsilon _{2}-\varepsilon _{1}\right) $ by Eq.(2.3.5).

If we want to continue the process and find the analogous properties to the
next Hamiltonian in the hierarchy, we first go through the intermediary
Hamiltonian which has zero energy for its ground state, and then repeat the
procedure outlined in detail above. The result is summarized in the Table 2.

\subsection{Constructing $\left( c_{n},d_{n}\right) $ from the spectrum of a
shape-invariant Hamiltonian}

We have just shown how to deduce the energy spectrum of a shape-invariant
Hamiltonian from knowledge of the parameters $\{c_n,d_n\}$. In this
subsection we will show the reverse: that from knowing the energy spectrum
of a shape-invariant Hamiltonian together with additional minimal
information we can create a tridiagonal representation of the
shape-invariant Hamiltonian.\newline
We recall Eq.(2.1.5) that $\varepsilon_1(\eta)=[c_n^2(\eta)-c_{n+1}^{2}]+[d_{n+1}^{2}-d_n^2]$
which is independent of the index $n$. Thus, choosing $n=0$, we also get 
$
\varepsilon_1(\eta)=[c_0^2(\eta)-c_{1}^{2}]+d_{1}^{2}
$
From the properties of the hierarchy of shape-invariant Hamiltonians
detailed in the previous subsection, we recognize that the relationship
between the set $\{c_n , d_n\}$ and $\varepsilon_1(\eta)$ is analogous to
the relationship between the set $\{c_{n+k},d_n\}$ and $\varepsilon_1(\eta+k%
\delta)$ . Therefore, we can write the analogous to Eq.(2.1.6), namely 
\begin{equation}
\varepsilon_1(\eta+k\delta)=[c_k^2(\eta)-c_{k+1}^{2}]+d_{1}^{2} \tag{2.6.1}
\end{equation}
Now we compare this with Eq.(2.1.5) when $n=k$. The result is 
\begin{equation}
\varepsilon_1(\eta)-\varepsilon_1(\eta+k%
\delta)=[d_{k+1}^2-d_{k}^{2}]-d_{1}^{2} \tag{2.6.2}
\end{equation}
We now sum each side of the above equation from $k=0$ to $k =m$ so as to
obtain 
\begin{equation}
(m+1)\varepsilon_1(\eta)-\sum\limits_{k=0}^{m}\varepsilon_1(\eta+k\delta)=%
\sum\limits_{k=0}^{m}[d_{k+1}^2-d_{k}^{2}]-(m+1)d_{1}^{2} \tag{2.6.3}
\end{equation}

Using the telescoping property to evaluate the right-hand sum, we get 
\begin{equation}
(m+1)\varepsilon_1(\eta)-\varepsilon_{m+1}(\eta)=d_{m+1}^2-(m+1)d_{1}^{2} \tag{2.6.4}
\end{equation}

Therefore, we have the major result that $d_{m+1}$, is determined by the
energy spectrum of the Hamiltonian and the parameter $d_1$. More
specifically, 
\begin{equation}
d_{m+1}^{2}=(m+1)d_1^2+[(m+1)\varepsilon_1(\eta)-\varepsilon_{m+1}(%
\eta)]. \tag{2.6.5}
\end{equation}
Furthermore, we already know from Eq.(2.3.6) that 
\begin{equation}
c_{m+1}^{2}(\eta)=c_{0}^{2}(\eta)+[(m+1)d_1-\varepsilon_{m+1}(\eta)]. \tag{2.6.6}
\end{equation}
From the above relations, we can construct a tridiagonal matrix
representation of a shape- invariant Hamiltonian from the details of its
energy spectrum together with the two parameters $\{c_{0}^{2}(\eta),d_1^2\}$.

\subsection{The superpotential for shape-invariant tridiagonal Hamiltonian}

In the 'differential' treatment of supersymmetry \cite{Cooper2001}, a quantity, called the
superpotential $W (x)$ , is introduced with intimate connection to the
ground state wave function, $\psi_0(x)$, of the quantum system as follows: 
\begin{equation}
W(x)=-\frac{(d/dx)\psi_0(x)}{\sqrt{2}\psi_0(x)} \tag{2.7.0}
\end{equation}
Although our treatment for the supersymmetry of tridiagonal Hamiltonians
does not need the concept of superpotential, yet we feel the need to show
that exist an analogous quantity having the expected property of the
superpotential. To do that, we first start from the basic quantities $A ,
A^{\dagger}$ and their associated representation in the basis. We then
define two related operators $\tilde{W}$ and $\tilde{D}$ as follows: 
\begin{equation}
\tilde{W}=\frac{1}{2}(A + A^{\dagger});\qquad \tilde{D}=\frac{1}{2}(A -
A^{\dagger}).  \label{W_tild,D_tild} \tag{2.7.1}
\end{equation}
Conversely, we have 
\begin{equation}
A=(\tilde{W}+\tilde{D});\qquad A^{\dagger}=(\tilde{W}-\tilde{D}).\tag{2.7.2}
\label{A,Adagger}
\end{equation}
We note that $\tilde{W}$ is hermitian while $\tilde{D}$ is antihermitian.
Also, 
\begin{eqnarray*}
H &=& A^{\dagger}A = (\tilde{W}-\tilde{D})(\tilde{W}+\tilde{D})=\tilde{W}^2-%
\tilde{D}\tilde{W}+\tilde{W}\tilde{D}-\tilde{D}^2\equiv H_0+V \\
H^{(+)} &=& AA^{\dagger} = (\tilde{W}+\tilde{D})(\tilde{W}-\tilde{D})=\tilde{%
W}^2+\tilde{D}\tilde{W}-\tilde{W}\tilde{D}-\tilde{D}^2\equiv H_0+V^{(+)}
\end{eqnarray*}

Comparing the form of the Hamiltonian and its supersymmetric partner, we can
identify the corresponding expressions as follows: 
\begin{equation}
H_0\rightarrow -\tilde{D}^2=\tilde{D}^{\dagger}\tilde{D}; \quad V
\rightarrow [\tilde{W}^2-\tilde{D}\tilde{W}+\tilde{W}\tilde{D}],\quad V^{(+)}
\rightarrow [\tilde{W}^2+\tilde{D}\tilde{W}-\tilde{W}\tilde{D}] \tag{2.7.3}
\end{equation}
This leads to the correspondence 
\begin{equation}
\tilde{D}\rightarrow \frac{1}{\sqrt{2}}\frac{d}{dx}. \tag{2.7.4}
\end{equation}
Taking this into account, and comparing relation Eq.(2.7.2) with the
following analogous relation in 'differential' supersymmetry 
\begin{equation}
A=(W+\frac{1}{\sqrt{2}}\frac{d}{dx});\qquad A^{\dagger}=(W-\frac{1}{\sqrt{2}}%
\frac{d}{dx}), \tag{2.7.5}
\end{equation}
we immediately make the correspondence 
\begin{equation}
\tilde{W}\rightarrow W \tag{2.7.6}
\end{equation}
Now we give several plausibility arguments to support the interpretation of $%
\tilde{W}$ as the superpotential. From the definition of $\tilde{W}$ in Eq.(2.7.1), we can write, 
\begin{equation}
\tilde{W}|\varepsilon_0\rangle=\frac{1}{2}(A+A^{\dagger})|\varepsilon_0%
\rangle=\frac{1}{2}(A^{\dagger})|\varepsilon_0\rangle \tag{2.7.7}
\end{equation}
since, as shown in Appendix C, the operator $A$ annihilates the ground
states $|\varepsilon _{0}\rangle $. This equivalent to 
\begin{equation}
\tilde{W}|\varepsilon _{0}\rangle =\frac{1}{2}(A^{\dagger }-A)|\varepsilon
_{0}\rangle =-\tilde{D}|\varepsilon _{0}\rangle . \tag{2.7.8}
\end{equation}%
More explicitly, if we assume that $\tilde{W}$ represent a local function of 
$x$, we then have 
\begin{equation}
\tilde{W}(x)\langle x|\varepsilon _{0}\rangle =-\langle x|\tilde{D}%
|\varepsilon _{0}\rangle . \tag{2.7.9}
\end{equation}%
We see that this form corresponds exactly to the form of Eq.(2.7.0).\newline
Additionally, we now show that $\tilde{W}$ related to the potentials $V\equiv V^{(-)}$ and 
$V^{(+)}$ as in the differential form of supersymmetry. We start with the
proposition that $V^{(\pm )}\rightarrow \tilde{W}^{2}\pm \tilde{D}\tilde{W}%
\mp \tilde{W}\tilde{D}$. Then for any state $|\varphi \rangle $, we have 
\begin{equation*}
V^{(\pm )}(x)\varphi (x) =\langle x|V^{(\pm )}|\varphi \rangle \rightarrow
\langle x|[\tilde{W}^{2}\pm \tilde{D}\tilde{W}\mp \tilde{W}\tilde{D}%
]|\varphi \rangle \tag{2.7.10}
\end{equation*}
\begin{equation*}
\qquad\qquad\qquad\qquad\qquad\qquad\qquad\qquad\,=\tilde{W}^{2}(x)\varphi (x)\pm \frac{1}{\sqrt{2}}\frac{d}{dx}[\tilde{W}%
(x)\varphi (x)]\mp \tilde{W}(x)\frac{1}{\sqrt{2}}\frac{d}{dx}[\varphi (x)] \tag{2.7.11} 
\end{equation*}
\begin{equation*}
\qquad=\tilde{W}^{2}(x)\varphi (x)\pm \frac{1}{\sqrt{2}}\frac{d}{dx}[\tilde{W}%
(x)]\varphi (x) \tag{2.7.12} 
\end{equation*}
\begin{equation*}
=\left[ \tilde{W}^{2}(x)\pm \frac{1}{\sqrt{2}}\frac{d\tilde{W}(x)}{dx}%
\right] \varphi (x) \tag{2.7.13}
\end{equation*}
Hence, 
\begin{equation}
V^{(\pm )}(x)=\tilde{W}^{2}(x)\pm \frac{1}{\sqrt{2}}\frac{d\tilde{W}(x)}{dx} \tag{2.7.14}
\end{equation}%
which is the result quoted in the literature if we identify $\tilde{W}$ as
the superpotential.\newline
Let's apply the above results to the Morse Hamiltonian. With 
\begin{equation}
\psi _{0}(x)=\sqrt{\frac{\alpha }{\Gamma (2D)}}\zeta ^{D}e^{-\zeta /2},\:\zeta=\zeta(x)=e^{-\alpha x}, \tag{2.7.15}
\end{equation}
and Eq(2.7.1) gives us 
\begin{equation}
W(x)=\frac{\alpha }{2\sqrt{2}}\left( 2D-\zeta (x)\right).\tag{2.7.16}
\end{equation}%
On the other hand, Eq.(2.7.9) says has that $\tilde{W}(x)=-\langle x|\widetilde{D}|\varepsilon_0\rangle=\frac{-1}{2}\langle x|(A-A^{\dagger})|\varepsilon_0\rangle$. Now $|\varepsilon _{0}\rangle$ is shown in appendix C to have the representation
\begin{equation}
|\varepsilon _{0}\rangle =\Lambda _{0}(0)\sum\limits_{n=0}^{\infty
}Q_{n}(0)|\phi _{n}\rangle \tag{2.7.17}
\end{equation}%
where 
\begin{equation}
Q_{n}(0)=\frac{(\gamma +\frac{1}{2}-D)_{n}}{\sqrt{n!(2\gamma +1)_{n}}},\;\: \Lambda _{0}(0)=\frac{\Gamma (\gamma +\frac{1}{2}+D)}{\sqrt{\Gamma
(2\gamma +1)\Gamma (2D)}}\tag{2.7.18}
\end{equation}
 Hence, 
\begin{equation}
\tilde{W}(x)\langle x|\varepsilon _{0}\rangle =-\frac{1}{2}\Lambda
_{0}(0)\sum\limits_{n=0}^{\infty }Q_{n}(0)\left[ d_{n}\langle x|\phi
_{n-1}\rangle -d_{n+1}\langle x|\phi _{n+1}\rangle \right]. \tag{2.7.19}
\end{equation}%
Detailed calculation gives 
\begin{equation}
\tilde{W}(x)\psi _{0}(x)=\frac{-\alpha \sqrt{\alpha }\Lambda _{0}(0)}{2\sqrt{%
2}\sqrt{\Gamma (2\gamma +1)}}\zeta ^{\gamma +\frac{1}{2}}e^{-\zeta
/2}\left[(\zeta -(2\gamma +1))-2\zeta \frac{d}{d\zeta }\right] F(\zeta ) \tag{2.7.20}
\end{equation}%
where $F(\zeta )=\sum\limits_{n=0}^{\infty }Q_{n}(0)L_{n}^{2\gamma }(\zeta )$%
. This sum is found to be $\zeta ^{D-(\gamma +\frac{1}{2})}\frac{\Gamma
(2\gamma +1)}{\Gamma (\gamma +\frac{1}{2}+D)}$. 
Now with 
\begin{equation*}
\psi _{0}(x)=\frac{\Lambda _{0}(0)\sqrt{\alpha \Gamma (2\gamma +1)}%
}{\Gamma (\gamma +\frac{1}{2}+D)}\zeta ^{D}e^{-\zeta /2},
\end{equation*}
we finally get 
\begin{equation}
\tilde{W}=\frac{\alpha }{2\sqrt{2}}(2D-\zeta ). \tag{2.7.21}
\end{equation}%
Which is identical to the result of Eq.(2.7.16).

\section{The coherent states associated with shape-invariant tridiagonal Hamiltonians}

We construct the coherent state $|z\rangle $ by displacing the ground state $%
|\varepsilon _{0}\rangle $ as 
\begin{equation*}
|z\rangle =\exp \left( zB^{\dagger }-z^{\ast }B\right) |\varepsilon
_{0}\rangle \tag{4.1}
\end{equation*}%
The exponential operator can be simplified by using the well-known
factorization \cite{Klauder}%
\begin{equation*}
\exp \left( zB^{\dagger }-z^{\ast }B\right) =e^{-\frac{1}{2}zz^{\ast }\left[
B,B^{\dagger }\right] }e^{zB^{\dagger }}e^{-z^{\ast }B} \tag{4.2}
\end{equation*}%
By also making use of the commutation relation of Eq.(2.1.10), we can
represent the coherent state in terms of the energy eigenstates of the
Hamiltonian as 
\begin{equation*}
|z\rangle =e^{-\frac{1}{2}zz^{\ast }\varepsilon _{1}\left( \eta -\delta
\right) }\sum\limits_{n=0}^{+\infty }\frac{z^{n}}{n!}\left( B^{\dagger
}\right) ^{n}|\varepsilon _{0}\rangle. \tag{4.6} 
\end{equation*}%
In subsection 2.2, we have explored how the raising operator acts on the
various eigenstates of the system. It follows from Eq.(2.2.0) that the
coherent state can be written as 
\begin{equation*}
|z\rangle =e^{-\frac{1}{2}zz^{\ast }\varepsilon _{1}\left( \eta -\delta
\right) }\sum\limits_{n=0}^{+\infty }\frac{z^{n}}{n!}\sqrt{%
\prod\limits_{j=0}^{n-1}\left( \sum\limits_{k=j}^{n-1}\varepsilon _{1}\left(
\eta +k\delta \right) \right) }|\varepsilon _{n}\rangle.  \tag{4.7} 
\end{equation*}%
We now show that these coherent states satisfy the resolution de the
identity operator the quantum Hilbert space $\mathcal{H}$ on which the Hamiltonian operator is acting. That is 
\begin{equation*}
1_{\mathcal{H}}=\int_{\mathbb{C}}|z\rangle \langle z| \varrho \left(
\left\vert z\right\vert ^{2}\right) \frac{d^{2}z}{\pi } \tag{4.8} 
\end{equation*}%
where $\varrho \left( \left\vert z\right\vert ^{2}\right) $ is an auxiliary
density function to be determined. By replacing $|z\rangle $ by its
expression in (4.7) inside the rank one operator $|z\rangle \langle z |$ and using polar coordinates $z=re^{i\theta },r>0,\theta
\in \left[ 0,2\pi \right) $ then Eq.(4.8) leads to 
\begin{equation*}
1=\int\limits_{0}^{+\infty }e^{-r^{2}\varepsilon _{1}\left( \eta -\delta
\right) }\frac{r^{2n+1}}{\left( n!\right) ^{2}}\prod\limits_{j=0}^{n-1}%
\left( \sum\limits_{k=j}^{n-1}\varepsilon _{1}\left( \eta +k\delta \right)
\right) \varrho \left( r^{2}\right) dr \tag{4.9} 
\end{equation*}%
More explicitly, we demand that the function $\varrho \left( r^{2}\right) $
\ solves the equation%
\begin{equation*}
\int\limits_{0}^{+\infty }e^{-r^{2}\varepsilon _{1}\left( \eta -\delta
\right) }r^{2n+1}\varrho \left( r^{2}\right) dr=\frac{\left( n!\right) ^{2}}{%
\prod\limits_{j=0}^{n-1}\left( \sum\limits_{k=j}^{n-1}\varepsilon _{1}\left(
\eta +k\delta \right) \right) }. \tag{4.10} 
\end{equation*}%
This is equivalent to the moment problem%
\begin{equation*}
\int\limits_{0}^{+\infty } x^n [e^{-x\varepsilon _{1}\left( \eta -\delta \right)
}\varrho \left( x\right)] dx=\frac{2\left( n!\right) ^{2}}{%
\prod\limits_{j=0}^{n-1}\left( \sum\limits_{k=j}^{n-1}\varepsilon _{1}\left(
\eta +k\delta \right) \right) } \tag{4.11} 
\end{equation*}%
for the function $x\longmapsto e^{-x\varepsilon _{1}\left( \eta -\delta \right) }\varrho
\left( x\right)$, which can be solved with the help of some known integral or by making an appeal to a transformation procedure, like Mellin \cite{Bateman} or Fourier. 

As an example, we consider the harmonic oscillator Hamiltonian for which $%
\varepsilon _{1}\left( \eta \right) =2\omega .$ Therefore, the quantity on
the right hand side of Eq.(4.11)  becomes $\varepsilon $%
\begin{equation*}
\int\limits_{0}^{\infty } x^n e^{-2\omega x}\varrho \left( x\right) dx=\frac{%
2\left( n!\right) ^{2}}{\left( 2\omega \right) ^{n}} \tag{4.12} 
\end{equation*}%
This is solved by the choice $\varrho \left( x\right) =4\omega .$

Another example is that of the Morse Hamiltonian where we have here $%
\varepsilon _{1}\left( \eta \right) =\frac{\alpha ^{2}}{2}\left( 2D-1\right) 
$ and $\eta =D,\delta =-1.$ Careful calculation shows that the quantity 
\begin{equation*}
\prod\limits_{j=0}^{n-1}\left( \sum\limits_{k=j}^{n-1}\varepsilon _{1}\left(
\eta +k\delta \right) \right) \tag{4.13}
\end{equation*}%
has the value 
\begin{equation*}
\prod\limits_{j=0}^{n-1}\left( \sum\limits_{k=j}^{n-1}\varepsilon _{1}\left(
\eta +k\delta \right) \right) =\left( \frac{\alpha ^{2}}{2}\right) ^{n}n!%
\frac{\Gamma \left( 2D-n +1\right) }{\Gamma \left( 2D-2n+1\right) }. \tag{4.14}
\end{equation*}%
This means that the function $\varrho \left( r^{2}\right) $ must satisfy the
moment relation%
\begin{equation*}
\int\limits_{0}^{+\infty }e^{-\frac{1}{2}\alpha ^{2}\left( 2D+1\right)
x}x^{n}\varrho \left( x\right) dx=\frac{2^{n+1}n!}{\alpha
^{2n}}\frac{\Gamma \left( 2D-2n +1\right) }{\Gamma \left( 2D-n +1\right) 
}. \tag{4.15}
\end{equation*}%
Of particular interest is the fact that the ground state is a coherent state 
$|z\rangle $ with $z=0$ since we have shown that $|0\rangle $ is just $%
|\varepsilon _{0}\rangle $. This is true in any defining scheme of
coherent states. One such scheme is the one that define \textit{"\`{a} la
Glauber"} the coherent state $|z\rangle $ as eigenvector of the
operator $B.$ That is $B|z\rangle =z|z\rangle $. Also, if we define $%
|z\rangle $ as the eignevector of the operator $A;$ i.e., $A|z\rangle
=z|z\rangle ,$ then $|0\rangle $ is just $|\varepsilon _{0}\rangle $. \ \
Now if we expand the vector in $|z\rangle $ in terms of the basis $\left\{
|\phi _{n}\rangle \right\} _{n=0}^{\infty }$ \ as 
\begin{equation*}
|z\rangle =\sum\limits_{n=0}^{+\infty }\Lambda _{n}\left( z\right) |\phi
_{n}\rangle  \tag{4.16}
\end{equation*}%
then the coefficients $\Lambda _{n}\left( z\right) $ can be found
recursively as $\Lambda _{n}\left( z\right) =\Lambda _{0}\left( z\right)
Q_{n}\left( z\right) $ where $Q_{0}\left( z\right) =1$ and 
\begin{equation*}
Q_{n}\left( z\right) =\prod\limits_{j=0}^{n-1}\left( \frac{z-c_{j}}{d_{j+1}}%
\right) ,\text{ }n\geq 1;\text{ \ \ }\Lambda _{0}\left( z\right) =\frac{1}{%
\sqrt{\sum_{n=0}^{+\infty }\left\vert Q_{n}\left( z\right) \right\vert ^{2}}}. \tag{4.17}
\end{equation*}%
This means that we can write the ground state vector in \ terms of the basis
as 
\begin{equation*}
|\varepsilon _{0}\rangle =\Lambda _{0}\left( 0\right)
\sum\limits_{n=0}^{+\infty }Q_{n}\left( 0\right) |\phi _{n}\rangle. \tag{4.18} 
\end{equation*}%
We conclude that the set of coefficients $\left( c_{n},d_{n}\right) $ and
the basis $\left\{ |\phi _{n}\rangle \right\} _{n=0}^{\infty }$ are
sufficient to write the ground state vector. \ We give two examples of such
construction. For the harmonic oscillator Hamiltonian \ using the relevant
quantities listed in Table 1, we easily find that 
\begin{equation*}
Q_{n}\left( 0\right) =\prod\limits_{j=0}^{n-1}\left( \frac{-c_{j}}{d_{j+1}}%
\right) =\left( \frac{\omega -\lambda ^{2}}{\omega +\lambda ^{2}}\right) ^{n}%
\sqrt{\frac{\Gamma \left( n+\nu +1\right) }{n!\Gamma \left( \nu +1\right) }}.\tag{4.19} 
\end{equation*}%
With the associated basis, given in Table 1, we have for $\psi _{0}\left( x\right)
=\langle x\mid \varepsilon _{0}\rangle $ :%
\begin{equation*}
\psi _{0}\left( x\right) =\Lambda _{0}\left( 0\right) x^{l+1}e^{-\frac{1}{2}%
x^{2}}\sqrt{\frac{2\lambda }{\Gamma \left( \nu +1\right) }}%
\sum\limits_{n=0}^{+\infty }t^{n}L_{n}^{\left( \nu \right) }\left(
x^{2}\right) ,\text{ \ }t=\left( \frac{\omega -\lambda ^{2}}{\omega +\lambda
^{2}}\right) .\tag{4.20} 
\end{equation*}%
But we know that for $\left\vert t\right\vert <1,$ which is satisfied here,
we have a closed form for the sum, namely%
\begin{equation*}
\sum\limits_{n=0}^{+\infty }t^{n}L_{n}^{\left( \nu \right) }\left( u\right)
=\left( 1-t\right) ^{-\nu -1}\exp \left( \frac{ut}{t-1}\right) .\tag{4.21} 
\end{equation*}%
Hence, we have 
\begin{equation*}
\psi _{0}\left( x\right) =\Lambda _{0}\left( 0\right) \sqrt{\frac{2\lambda }{%
\Gamma \left( \nu +1\right) }}\left( \frac{\omega +\lambda ^{2}}{2\lambda
^{2}}\right) ^{\nu +1}x^{l+1}e^{-\frac{1}{2}\omega x^{2}}.\tag{4.22} 
\end{equation*}%
On the other hand, we have 
\begin{equation*}
\left( \Lambda _{0}\left( 0\right) \right) ^{-2}=\sum\limits_{n=0}^{+\infty
}\left\vert Q_{n}\left( 0\right) \right\vert ^{2}=\sum\limits_{n=0}^{+\infty
}L_{n}^{\left( \nu \right) }\left( 0\right) \left( t^{2}\right) ^{n}=\left(
1-t^{2}\right) ^{-\nu -1}\tag{4.23} 
\end{equation*}%
\begin{equation*}
=\left( \frac{2\lambda ^{2}}{\omega +\lambda ^{2}}\right) ^{-\nu -1}\left( 
\frac{2\omega }{\omega +\lambda ^{2}}\right) ^{-\nu -1}.\tag{4.24} 
\end{equation*}%
Thus, we finally have 
\begin{equation*}
\psi _{0}\left( x\right) =\sqrt{\frac{2\sqrt{\omega }}{\Gamma \left( \nu
+1\right) }}\left( \sqrt{\omega }r\right) ^{l+1}e^{-\frac{1}{2}\omega x^{2}}.\tag{4.25} 
\end{equation*}%
which is indeed the ground state wave function. It is important to notice
that the wave function is, as expected, independent of the free scale parameter 
$\lambda $ which characterizes the basis not the physical system. We give the
details for the ground state of the Morse Hamiltonian in Appendix D.

\section{Concluding remarks}

We would like to emphasize that several important points regarding the
supersymmetric properties of shape-invariant tridiagonal Hamiltonians.
First, the basic quantities in our approach are the parameters $\left(
c_{n},d_{n}\right) $ which are related to matrix elements of the tridiagonal
Hamitonian by Eq.$\left( 1.3\right)$-Eq.$\left( 1.4\right)$. Second, for Hamiltonians with
shape-invariance property, an important derived quantity is the energy of
the first excited state $\varepsilon _{1}\left( \eta \right) $ which plays a
pivotal role in the complete description of the Hamiltonian energy spectrum
and that of its supersymmetric partner. Third, the approach adopted here
accomodates familiar concepts such as the superpotential although they play
no role in the analysis . Finally, the need is evident to catalogue as many
cases of Hamiltonians having tridiagonal matrix representation in specific
basis. The effort of Alhaidari \textit{et al} \cite{Alhaidari} in this
regard is commendable.\\
\\
\textbf{{\large Appendix A.}} For the case $n =0$, consider the vector $B^{(+)}|\varepsilon _{0}\rangle $%
. Now 
\begin{eqnarray*}
&(&B^{(+)}B)[B^{(+)}|\varepsilon_0\rangle]=B^{(+)}(BB^{(+)})|\varepsilon_0\rangle=B^{(+)}[B^{(+)}B+\varepsilon_1(\eta-\delta)]|\varepsilon_0\rangle\cr
&=& B^{(+)}\varepsilon_1(\eta-\delta)|\varepsilon_0\rangle=\varepsilon_1(\eta)[B^{(+)}|\varepsilon_0\rangle]
\end{eqnarray*}%
since $(B^{(+)}B)|\varepsilon _{0}\rangle =(A^{(+)}A)|\varepsilon
_{0}\rangle =H|\varepsilon _{0}\rangle =0$ and by using Eq.(2.1.10). This
means that the vector $[B^{(+)}|\varepsilon _{0}\rangle ]$ is proportional
to the energy eigenvector $|\varepsilon _{1}\rangle $ with eigenvalue $%
\varepsilon _{1}(\eta )$. Hence $[B^{(+)}|\varepsilon _{0}\rangle ]\infty
|\varepsilon _{1}(\eta )\rangle $.\newline
The normalized version of this vector is 
\begin{equation}
|\varepsilon _{1}(\eta )\rangle =\frac{1}{\sqrt{|\varepsilon _{1}(\eta
)\rangle }}B^{(+)}|\varepsilon _{0}\rangle .\tag{A.1}
\end{equation}%
To see this, note that 
\begin{eqnarray*}
&\langle&\varepsilon_1(\eta)|\varepsilon_1(\eta)\rangle=\langle\varepsilon_0|B \frac{1}{\sqrt{|\varepsilon_1(\eta)\rangle}}\frac{1}{\sqrt{|\varepsilon_1(\eta)\rangle}}B^{(+)}|\varepsilon_0\rangle\cr
&=&\langle\varepsilon_0|B \frac{1}{\sqrt{|\varepsilon_1(\eta)\rangle}}B^{(+)}|\varepsilon_0\rangle=\langle\varepsilon_0 B B^{(+)}\frac{1}{\varepsilon_1(\eta-\delta)}|\varepsilon_0\rangle\cr
&=& \langle\varepsilon_0| [B^{(+)}B+\varepsilon_1(\eta-\delta)]\frac{1}{\varepsilon_1(\eta-\delta)}|\varepsilon_0\rangle=1+\langle \varepsilon_0[A^{(+)}A]\frac{1}{\varepsilon_1(\eta-\delta)}|\varepsilon_0\rangle=1
\end{eqnarray*}%
We now assume that the result holds for the case $k \leq n $. We then
consider the right-hand side of Eq.(2.2.0) for $n+1$. We use the
induction hypothesis to get : 
\begin{eqnarray*}
&\frac{1}{\sqrt{\displaystyle\sum_{k=0}^{n+1}\varepsilon_1(\eta+k\delta)}}&B^{(+)}\frac{1}{\sqrt{\displaystyle\sum_0^{n}\varepsilon_1(\eta+k\delta)}}B^{(+)}\cdots\frac{1}{\sqrt{\varepsilon_1(\eta)+\varepsilon_1(\eta+\delta)}}B^{(+)}\frac{1}{\sqrt{\varepsilon_1(\eta)}}B^{(+)}|\varepsilon_0\rangle\cr
&=&\frac{1}{\sqrt{\displaystyle\sum_{k=0}^{n+1}\varepsilon_1(\eta+k\delta)}}B^{(+)}|\varepsilon_{n+1}\rangle.
\end{eqnarray*}%
Now wee proceed to show that the resulting vector $\frac{1}{\sqrt{%
\displaystyle\sum_{0}^{n +1}\varepsilon _{1}(\eta +k\delta )}}%
B^{(+)}|\varepsilon _{n +1}\rangle $ is an eigenvector of the Hamiltonian
with eigenvalue $\varepsilon _{n +2}$. Specifically, we have 
\begin{eqnarray*}
&H&[\frac{1}{\sqrt{\displaystyle\sum_{k=0}^{n+1}\varepsilon_1(\eta+k\delta)}}B^{(+)}|\varepsilon_{n+1}\rangle]=\frac{1}{\sqrt{\displaystyle\sum_0^{n+1}\varepsilon_1(\eta+k\delta)}}(A^{(+)}A)B^{(+)}|\varepsilon_{n+1}\rangle\cr
&=& \frac{1}{\sqrt{\displaystyle\sum_{k=0}^{n+1}\varepsilon_1(\eta+k\delta)}}(B^{(+)}B)B^{(+)}|\varepsilon_{n+1}\rangle=\varepsilon_{n+1} B^{(+)}(B^{(+)}B)|\varepsilon_{n+1}\rangle\cr
&=&  \frac{1}{\sqrt{\displaystyle\sum_0^{n+1}\varepsilon_1(\eta+k\delta)}}B^{(+)}[B^{(+)}B+\varepsilon_1(\eta-\delta)]|\varepsilon_{n+1}\rangle \cr
&=&\frac{1}{\sqrt{\displaystyle\sum_{k=0}^{n+1}\varepsilon_1(\eta+k\delta)}}B^{(+)}[\varepsilon_{n+1}(\eta)+\varepsilon_1(\eta-\delta)]|\varepsilon_{n+1}\rangle
\end{eqnarray*}%
But, 
\begin{equation*}
\varepsilon _{n +1}(\eta )+\varepsilon _{1}(\eta -\delta )=\displaystyle%
\sum_{k=0}^{n +1}\varepsilon _{1}(\eta +k\delta )\varepsilon (\eta -\delta
)=\displaystyle\sum_{k=-1}^{n +1}\varepsilon _{1}(\eta +k\delta )
\end{equation*}%
Therefore, 
\begin{eqnarray*}
H[\frac{1}{\sqrt{\displaystyle\sum_{k=0}^{n +1}\varepsilon _{1}(\eta
+k\delta )}}B^{(+)}|\varepsilon _{n +1}\rangle ]=\frac{1}{\sqrt{%
\displaystyle\sum_{k=0}^{n +1}\varepsilon _{1}(\eta +k\delta )}}B^{(+)}[%
\displaystyle\sum_{k=-1}^{n +1}\varepsilon _{1}(\eta +k\delta
)]|\varepsilon _{n +1}\rangle \cr=\frac{[\displaystyle\sum_{k=-1}^{n
+1}\varepsilon _{1}(\eta +(k+1)\delta )]}{\sqrt{\displaystyle\sum_{k=0}^{n
+1}\varepsilon _{1}(\eta +k\delta )}}B^{(+)}|\varepsilon _{n +1}\rangle =%
\frac{\displaystyle\sum_{j=0}^{n +2}\varepsilon _{1}(\eta +j\delta )}{%
\sqrt{\displaystyle\sum_{k=0}^{n +1}\varepsilon _{1}(\eta +k\delta )}}%
B^{(+)}|\varepsilon _{n +1}\rangle \cr=\varepsilon _{n +2}[\frac{1}{%
\sqrt{\displaystyle\sum_{k=0}^{n +1}\varepsilon _{1}(\eta +k\delta )}}%
B^{(+)}|\varepsilon _{n +1}\rangle ]
\end{eqnarray*}%
It is easy to show that the vector $\left[\sqrt{\displaystyle%
\sum_{k=0}^{n +1}\varepsilon _{1}(\eta +k\delta )}\right]^{-1}B^{(+)}|\varepsilon
_{n +1}\rangle $ is indeed normalized. This completes the proof. 
\newline
\textbf{{\large Appendix B.}} In this Appendix, we show in the case of the tridiagonal harmonic oscillator
Hamiltonian, that $B=TA,\,(T=e^{-\delta \partial /\partial \eta })$ yields a
lowering operator associated with the harmonic oscillator. We recall the
signature property is that this operator has the following effect on the
energy eigenstates: 
\begin{equation*}
B|\varepsilon _{m }\rangle =\sqrt{\varepsilon _{m }}|\varepsilon _{m
-1}\rangle . \tag{B.1}
\end{equation*}%
In particular, it annihilates the ground state since $\varepsilon _{0}=0$.
In fact, since $\varepsilon _{m }=2m \omega $, we should be able to show that, for the
harmonic oscillator, 
\begin{equation*}
B|\varepsilon _{m }\rangle =\sqrt{2m \omega }|\varepsilon _{m
-1}\rangle . \tag{B.2}
\end{equation*}%
We recall from Table 1 that the basis that makes the harmonic oscillator
tridiagonal is given by
\begin{equation*}
\phi _{n}(r)=\langle r|\phi _{n}\rangle =\sqrt{\frac{2\lambda n!}{\Gamma
(n+\nu+1)}}(\lambda r)^{l+1}e^{-\lambda ^{2}r^{2}/2}L_{n}^{(\nu )}(\lambda
^{2}r^{2}),\,\nu \equiv l+\frac{1}{2} \tag{B.3}
\end{equation*}%
On the other hand, the energy eigenstate has the explicit form: 
\begin{equation*}
\chi _{n}(r)=\langle r|\varepsilon _{m }\rangle =\sqrt{\frac{2\sqrt{\omega 
}m !}{\Gamma (n+v+1)}}(\sqrt{\omega }r)^{l+1}e^{-\omega r^{2}/2}L_{n}^{(\nu )
}(\omega r^{2}),\,\nu \equiv l+\frac{1}{2}.\tag{B.4}
\end{equation*}%
If we now write 
\begin{equation*}
|\varepsilon _{m }\rangle =\displaystyle\sum_{n=0}^{+\infty }|\phi
_{n}\rangle \Gamma _{n,m },\tag{B.4}
\end{equation*}%
then $\Gamma _{n,m }=\langle \phi _{n}|\varepsilon _{m }\rangle =%
\displaystyle\int_{0}^{\infty }\phi _{n}(r)\chi _{m }(r)$. Detailed
performance of this integral yields the following explicit result  \medskip\\
\begin{equation*}
\Gamma_{n,m}=\sqrt{\frac{4\lambda \sqrt{\omega} n!m!}{\Gamma(n+\nu+1)\Gamma(m+\nu+1)}}\frac{(\lambda \sqrt{\omega})^{l+1}}{2}\frac{\Gamma(n+m+\nu+1)}{n!m!} \tag{B.5}
\end{equation*}
\begin{equation*}
\qquad \qquad\times \,
\frac{(-1)^{m}(\frac{\omega}{2}-\frac{\lambda^2}{2})^{n+m}}{(\frac{\omega}{2}+\frac{\lambda^2}{2})^{n+m+\nu+1}}\, _2F_1\left(-m,-n;-(n+m+\nu);\left(\frac{\lambda^2+\omega}{\lambda^2-\omega}\right)^2\right). \tag{B.6}
\end{equation*}
With $B=TA,\,(T=e^{-\delta \partial /\partial \eta })$, together with the
expansion of the energy eigenstate in terms of the basis, we have:
\begin{equation*}
B|\varepsilon _{m }\rangle =e^{-\delta \partial /\partial \eta
}\sum_{n=0}^{+\infty }A|\phi _{n}\rangle \Gamma _{n,m }.\tag{B.7}
\end{equation*}%
But we know that $A|\phi _{n}\rangle =c_{n}|\phi _{n}\rangle +d_{n}|\phi
_{n-1}\rangle $. Hence, 
\begin{equation*}
B|\varepsilon _{m }\rangle = e^{-\delta \partial /\partial \eta
}\sum_{n=0}^{+\infty }\{c_{n}|\phi _{n}\rangle +d_{n}|\phi _{n-1}\rangle
\}\Gamma _{n,m }
=e^{-\delta \partial /\partial \eta
}\sum_{n=0}^{+\infty }\{c_{n}(\eta )\Gamma _{n,m }(\eta )+d_{n+1}\Gamma
_{n+1,m }(\eta )\}|\phi _{n}\rangle \tag{B.8}
\end{equation*}
Using Eq.(B.6), we can write explicitly the quantity in parenthesis as 
\begin{eqnarray*}
\{c_{n}(\eta )\Gamma _{n,m }(\eta )+d_{n+1}\Gamma _{n+1,m }(\eta )\}=%
\frac{\sqrt{2}}{\lambda }\sqrt{\frac{4\lambda \sqrt{\omega }n!m !}{\Gamma
(n+\nu +2)\Gamma (m +\nu +1)}}\frac{(\lambda \sqrt{\omega })^{l+1}}{2}%
\frac{\Gamma (n+m +\nu +1)}{n!m !}\cr\frac{(-1)^{m }(\frac{\omega }{2}-%
\frac{\lambda ^{2}}{2})^{n+m }}{(\frac{\omega }{2}+\frac{\lambda ^{2}}{2}%
)^{n+m +\nu +1}}\left[ -(n+m +\nu )\,_{2}F_{1}\left( -m ,-(m +\nu
);-(n+m +\nu );\frac{\tau }{\tau }\right) \right. 
\end{eqnarray*}
\begin{equation*}
 +\left. (n+m +\nu
+1)\,_{2}F_{1}\left( -m ,-(m +\nu );-(n+m +\nu )-1;\frac{\tau }{\tau }%
\right) \right] \tag{B.9}
\end{equation*}
where we have used the abbreviation $\tau =\left( \frac{\lambda ^{2}+\omega 
}{\lambda ^{2}-\omega }\right) ^{2}$ and the relation (\cite{MOS},p.47 )
\begin{equation*}
\,_{2}F_{1}(a,b;c;z)=(1-z)^{-a}\,_{2}F_{1}\left(a,c-b;c;\frac{z}{z-1}\right)\tag{B.10}
\end{equation*}%
Now we use further the relation (\cite{MOS},p.46 )
\begin{equation*}
(1-c)\,_{2}F_{1}(a,b;c-1;z)+(c-a-1)\,_{2}F_{1}(a,b;c;z)+a%
\,_{2}F_{1}(a+1,b;c;z)=0\tag{B.11}
\end{equation*}
for parameters
\begin{equation*}
a=-m ,\, b=-(m +\nu )\,\, \text{and}\,\,c=-(n+m +\nu ).
\end{equation*}%
Therefore $c_{n}(\eta )\Gamma _{n,m }(\eta )+d_{n+1}\Gamma _{n+1,m }(\eta )=%
\sqrt{2m \omega }\Gamma _{n,m -1}(\eta +1);\quad \eta \equiv \nu $. With $\delta =+1$, Eq.(B.8) becomes 
\begin{equation*}
B|\varepsilon _{m }\rangle =\sqrt{2m \omega }\sum_{n=0}^{+\infty }\Gamma
_{n,m -1}(\eta )|\phi _{n}\rangle =\sqrt{2m \omega }|\varepsilon _{m
-1}\rangle =\sqrt{\varepsilon _{m }}|\varepsilon _{m -1}\rangle \tag{B.12}
\end{equation*}%
This shows that the operator $B=TA,\,(T=e^{-\delta \partial /\partial \eta })$
is the lowering operator for the Harmonic oscillator Hamiltonian.
\newline\\
\textbf{{\large Appendix C.}} We can check that $B|\varepsilon _{0}\rangle =0$ as follows. \ Using the
expansion of $|\varepsilon _{0}\rangle $
\begin{equation*}
|\varepsilon _{0}\rangle =\sqrt{\Omega (\varepsilon _{0})}%
\sum\limits_{n=0}^{\infty }P_{n}(\varepsilon _{0})|\phi _{n}\rangle  \tag{4.3}
\end{equation*}%
then, the action of the operator $A$ on this state is 
\begin{equation*}
A|\varepsilon _{0}\rangle  =\sqrt{\Omega (\varepsilon _{0})}%
\sum\limits_{n=0}^{\infty }P_{n}(\varepsilon _{0})\left( c_{n}|\phi _{n}\rangle
+d_{n}|\phi _{n-1}\rangle \right)  \tag{4.4} 
\end{equation*}
\begin{equation*}
\qquad\qquad=\sqrt{\Omega (\varepsilon _{0})}\sum\limits_{n=0}^{\infty
}\left( p_{n}(\varepsilon _{0})c_{n}+P_{n+1}(\varepsilon _{0})d_{n+1}\right)|\phi
_{n}\rangle \tag{4.5}
\end{equation*}
Since $c_{n}^{2}=-b_{n}\frac{P_{n+1}(\varepsilon _{0})}{P_{n}(\varepsilon
_{0})}$ and $d_{n}^{2}=-b_{n}\frac{P_{n}(\varepsilon _{0})}{%
P_{n+1}(\varepsilon _{0})}$, it follows that $A|\varepsilon _{0}\rangle =0$.
Recalling that $B=TA$, it also follows that $B|\varepsilon _{0}\rangle =0$.
\newline\\
\textbf{{\large Appendix D.}} The ground state ground state is a coherent state $|z\rangle $ with $z=0$.
The eigenvector can be written in terms of the basis as 
\begin{equation}
|\varepsilon _{0}\rangle =\Lambda _{0}(0)\sum\limits_{n=0}^{\infty
}Q_{n}(0)|\phi _{n}\rangle ,\tag{C.1}
\end{equation}%
where $Q_{0}(z)=1,\,Q_{n}(z)=\prod\limits_{j=0}^{n-1}\left( \frac{z-c_{j}}{%
d_{j+1}}\right) $ for $n\geq 1$, and $\Lambda _{0}(z)=\frac{1}{\sqrt{%
\sum\limits_{n=0}^{\infty }|Q_{n}(z)|^{2}}}$.\newline
Here we give details of the calculation of the ground state of the Morse
Hamiltonian using the relevant parameter in Table l. We can easily find that 
\begin{equation*}
Q_{n}(0)=\frac{\Gamma (n+\gamma +\frac{1}{2}-D)}{\Gamma (\gamma +\frac{1}{2}%
-D)}\sqrt{\frac{\Gamma (2\gamma +1)}{n!\Gamma (n+2\gamma +1)}}=\frac{(\gamma
+\frac{1}{2}-D)_{n}}{\sqrt{n!(2\gamma +1)_{n}}},\quad  n=0,1,2, \cdots .\tag{C.2}
\end{equation*}%
With the basis that renders the Morse Hamiltonian tridiagonal, the ground
state wave function has the form 
\begin{equation*}
\psi _{0}(\zeta )=\Lambda _{0}(0)\sqrt{\frac{\alpha }{\Gamma (2\gamma +1)}}%
\zeta ^{\gamma +\frac{1}{2}}e^{-\zeta /2}\sum\limits_{n=0}^{\infty }\frac{%
(\gamma +\frac{1}{2}-D)_{n}}{(2\gamma +1)_{n}}L_{n}^{2\gamma }(\zeta )\tag{C.3}
\end{equation*}%
On the one hand, 
\begin{equation*}
\sum\limits_{n=0}^{\infty }\frac{(\gamma +\frac{1}{2}-D)_{n}}{(2\gamma
+1)_{n}}L_{n}^{2\gamma }(\zeta ) =\lim\limits_{t\rightarrow
1^{-}}\sum\limits_{n=0}^{\infty }\frac{(\gamma +\frac{1}{2}-D)_{n}}{(2\gamma
+1)_{n}}L_{n}^{2\gamma }(\zeta )t^{n} \tag{C.4}
\end{equation*}
\begin{equation*}
\qquad\qquad\qquad\qquad\qquad\qquad\qquad\qquad\qquad\qquad\quad=\lim\limits_{t\rightarrow 1^{-}}(1-t)^{-(\gamma +\frac{1}{2}%
-D)}{}_{1}F_{1}\left( \gamma +\frac{1}{2}-D;\,2\gamma +1;\,\frac{-t\zeta }{%
1-t}\right) .
\end{equation*}%
Now we make use of two important relations (\cite{MOS}, p.267): 
\begin{equation}
{}_{1}F_{1}(a;b;z)=e^{z}{}_{1}F_{1}(b-a;b;-z)\tag{C.5}
\end{equation}%
and (\cite{MOS}, p.289) 
\begin{equation}
{}_{1}F_{1}\left( a,\,c;z\right) \sim \frac{\Gamma (c)}{\Gamma (a)}%
e^{z}z^{a-c}\quad \text{as}\quad Re(z)\rightarrow +\infty\tag{C.6}
\end{equation}%
This means that 
\begin{equation}
\sum\limits_{n=0}^{\infty }\frac{(\gamma +\frac{1}{2}-D)_{n}}{(2\gamma
+1)_{n}}L_{n}^{2\gamma }(\zeta )=\frac{\Gamma (2\gamma +1)}{\Gamma (\gamma +%
\frac{1}{2}+D)}\zeta ^{-(\gamma +\frac{1}{2})+D}.\tag{C.7}
\end{equation}%
On the other hand, 
\begin{equation}
\sum\limits_{n=0}^{\infty }|Q_{n}(0)|^{2}=\sum\limits_{n=0}^{\infty }\frac{%
(\gamma +\frac{1}{2}-D)_{n}(\gamma +\frac{1}{2}-D)_{n}}{n!(2\gamma +1)_{n}}%
={}_{2}F_{1}\left( \gamma +\frac{1}{2}-D,\,\gamma +\frac{1}{2}-D;\,2\gamma
+1;\,1\right) .\tag{C.8}
\end{equation}%
But (\cite{MOS}, p.40) :
\begin{equation}
{}_{2}F_{1}(a,b;c;1)=\frac{\Gamma (c)\Gamma (c-a-b)}{\Gamma (c-a)\Gamma (c-b)%
}, \qquad Re(a+b-c)<0,\, c\neq 0,-1,-2,\cdots .\tag{C.9}
\end{equation}%
Therefore 
\begin{equation}
\sum\limits_{n=0}^{\infty }|Q_{n}(0)|^{2}=\frac{\Gamma (2\gamma +1)\Gamma
(2D)}{\Gamma (\gamma +\frac{1}{2}+D)\Gamma (\gamma +\frac{1}{2}+D)}.\tag{C.10}
\end{equation}%
Thus, 
\begin{equation}
\Lambda _{0}(0)=\frac{1}{\sqrt{\sum\limits_{n=0}^{\infty }|Q_{n}(0)|^{2}}}=%
\frac{\Gamma (\gamma +\frac{1}{2}+D)}{\sqrt{\Gamma (2\gamma +1)\Gamma (2D)}}.\tag{C.11}
\end{equation}%
Combining all above results together, we finally get the following explicit
form of the ground state wave function 
\begin{equation*}
\psi _{0}(\zeta )=\sqrt{\frac{\alpha }{\Gamma (2D)}}\zeta ^{D}e^{-\zeta /2}.\tag{C.12}
\end{equation*}%
It is to be noticed, again, that the wave function is independent of the
free scale parameter $\gamma $ which characterizes the chosen basis.
\\ \\ 
\textbf{Table 1.} Parameters and results K.E., radial oscillator and Morse hamiltonians.
\\
\begin{small}
\begin{tabular}{|c|c|c|c|}
\hline
 & $K.$ $E$ & $Harmonic$ $Oscillator$ & $Morse$ $Oscillator$ \\ \hline
$H$ & $-\frac{1}{2}\frac{d^2}{dr^2}+\frac{l(l+1)}{2r^2}$ &$ -\frac{1}{2}\frac{d^2}{dr^2}+\frac{l(l+1)}{2r^2}+\frac{1}{2}\omega^2 r^2-(l+\frac{3}{2})\omega$ & $-\frac{1}{2}\frac{d^2}{dx^2}+V_0(e^{-2\alpha x} -2e^{-\alpha x})+\frac{1}{2} \alpha^2 D^2$ \\ \hline
$\phi_n(y)$ & $K_n(y)^{\nu +\frac{1}{2}}e^{-\frac{y^2}{2}}L_n^{\nu}(y^2)$ &
$K_n(y)^{\nu +1/2}e^{-y^2/2}L_n^{\nu}(y^2)$ & $K_n(y)^{\gamma +1/2}e^{-y/2}L_n^{(\gamma )}(y)$
\\ \hline
$y$& $\lambda r$ & $\lambda r$ & $\frac{\sqrt{8V_0}}{\alpha}e^{-\alpha x}$\\ \hline
$K_n$ & $\sqrt{\frac{(2\lambda)n!}{\Gamma(n+\nu+1)}}$ & $\sqrt{\frac{(2\lambda)n!}{\Gamma(n+\nu+1)}}$ & $\sqrt{\frac{(\alpha)n!}{\Gamma(n+2\gamma+1)}}$ \\ \hline
Free parameters & $\lambda$ & $\lambda$ & $\gamma$ \\ \hline
Other parameter & $\nu=l+\frac{1}{2}$ & $\nu=l+\frac{1}{2}$ & $D=\frac{\sqrt{2V_0}}{\alpha}-\frac{1}{2}$
\\ \hline
$c_{n}^{2}$ & $\frac{\lambda ^{2}}{2}(n+\nu +1)$ & $\frac{(\lambda -\frac{%
\omega }{\lambda })^{2}}{2}(n+\nu +1)$ & $\frac{\alpha ^{2}}{2}\left(
n+\gamma +\frac{1}{2}-D\right) ^{2}$ \\ \hline
$d_{n}^{2}$ & $\frac{\lambda ^{2}}{2}n$ & $\frac{(\lambda +\omega /\lambda )%
}{2}n$ & $\frac{\alpha ^{2}}{2}n(n+2\gamma )$ \\ \hline
$\varepsilon _{1}(\eta )$ & $0$ & $2\omega $ & $\frac{\alpha ^{2}}{2}(2D-1)$
\\ \hline
$\eta $ & $\nu $ & $\nu $ & $D$
 \\ \hline
$\delta $ & $1$ & $1$ & $-1$
 \\ \hline
$\varepsilon _{1}(\eta -\delta )$ & $0$ & $2\omega $ & $\frac{\alpha ^{2}}{2%
}(2D+1)$ 
 \\ \hline
$\varepsilon _{\mu }^{(+)}(\eta )$ & $0$ & $2(\mu +1)\omega $ & $\frac{%
\alpha ^{2}}{2}(\mu +1)(2D-(\mu +1))$
\\ \hline
$\lbrack B,B^{\dag }]$ & $0$ & $2\omega $ & $\frac{\alpha ^{2}}{2}(2D-1)$ 
\\ 
\hline
$V(r)$ & $\frac{l(l+1)}{2r^{2}}$ & $\frac{l(l+1)}{2r^{2}}+\frac{1}{2}\omega
^{2}r^{2}-\left( l+\frac{3}{2}\right) \omega $ & $\frac{\alpha ^{2}}{2}%
\left( D+\frac{1}{2}\right)^2(e^{-2\alpha x} -2e^{-\alpha x})+
\frac{\alpha ^{2}}{2}%
D^2 $ \\ \hline
$V^{(+)}(r)$ & $\frac{(l+1)(l+2)}{2r^{2}}$ & $\frac{(l+1)(l+2)}{2r^{2}}+%
\frac{1}{2}\omega ^{2}r^{2}-\left( l+\frac{3}{2}\right) \omega $ & $\frac{\alpha ^{2}}{2}%
\left( D-\frac{1}{2}\right)^2(e^{-2\alpha x} -2e^{-\alpha x})+
\frac{\alpha ^{2}}{2}%
D^2 $ \\ \hline
\end{tabular}
\end{small}
\\
\\
\\
\textbf{Table 2.} Basic parameters for the hierarchy of supersymmetric partner hamiltonians.
\medskip
\\
\bigskip
\begin{tabular}{|c|c|c|c|c|c|}
\hline
The Hamiltonian & Ground state energy & $c_n$ & $d_n$ & $a_n$ & $b_n$
\\ 
\hline
$H$ & 0 & $c_n$ & $d_n$ & $c_n^2+d_n^2$ & $c_n d_{n+1}$ \\ 
\hline
$H^{(+)}$ & $\varepsilon_1(\eta)$ & $d_{n+1}$ & $c_n$ & $c_n^2+d_{n+1}^2$ & $c_{n+1}d_{n+1}$ \\ 
\hline
$\tilde{H}^{(+)}=H^{(+)}-\varepsilon_1(\eta)$ & $0$ & $c_{n+1}$ & $d_n$ & $c_{n+1}^2+d_{n}^2$ & $c_{n+1}d_{n+1}$ \\ 
\hline
$H^{(++)}$ & $\varepsilon_1(\eta+\delta)$ & $d_{n+1}$ & $c_{n+1}$ & $c_{n+1}^2+ d_{n+1}^2$ & $c_{n+2}d_{n+1}$ \\ 
\hline
$\tilde{H}^{(++)}=H^{(++)}-\varepsilon_1(\eta+\delta)$ & $0$ & $c_{n+2}$  & $d_n$ & $c_{n+2}^2+d_n^2$ & $c_{n+2}d_{n+1}$  \\ 
\hline
$H^{(+++)}$ & $\varepsilon_1(\eta+2\delta)$ & $d_{n+1}$ & $c_{n+2}$ & $c_{n+2}^2+d_{n+1}^2$& $c_{n+3}d_{n+1}$ \\ 
\hline
....& ....&....&....&....&....\\ 
\hline
$\widetilde{H}^{k(+)}$ & $0$ & $c_{n+k}$ & $d_n$ & $c_{n+k}^2+d_{n}^2$ & $c_{n+k}d_{n+1}$ \\ 
\hline
$H^{(k+1)(+)}$ & $\varepsilon_1(\eta+k\delta)$ & $d_{n+1}$ & $c_{n+k}$ & $c_{n+k}^2+d_{n+1}^2$ & $c_{n+k+1}d_{n+1}$\\
\hline
\end{tabular}

\begin{footnotesize}

\end{footnotesize}

\end{document}